\documentclass[aps,prl,twocolumn,showpacs,superscriptaddress]{revtex4}

\usepackage{graphicx}
\usepackage{dcolumn}
\usepackage{bm}

\begin{document}

\preprint{submitted to PRL}

\title{Genuine Electronic States of 
Vanadium Perovskites Revealed by High-Energy Photoemission}


\author{A. Sekiyama}
\email[]{sekiyama@mp.es.osaka-u.ac.jp}
\author{H. Fujiwara}
\author{S. Imada}
\affiliation{Department of Material Physics, 
Graduate School of Engineering Science, Osaka 
University, Toyonaka, Osaka 560-8531, Japan}
\author{H. Eisaki}
\thanks{Present Address: National Institute of Advanced Industrial 
Science and Technology (AIST), Umezono, Tsukuba, Ibaraki 305-8568, 
Japan}
\author{S. I. Uchida}
\affiliation{Department of Advanced Materials Science, 
Graduate School of Frontier Sciences, 
University of Tokyo, Tokyo 113-8656, Japan}
\author{K. Takegahara}
\affiliation{Department of Materials Science and Technology, 
Hirosaki University, Hirosaki, Aomori 036-8561, Japan}
\author{H. Harima}
\affiliation{The Institute of Scientific and Industrial Research, 
Osaka University, Ibaraki, Osaka 567-0047, Japan}
\author{Y. Saitoh}
\affiliation{Department of Synchrotron Research, Kansai Research 
Establishment, Japan Atomic Energy Research Institute, SPring-8, 
Mikazuki, Hyogo 679-5198, Japan}
\author{S. Suga}
\affiliation{Department of Material Physics, 
Graduate School of Engineering Science, Osaka 
University, Toyonaka, Osaka 560-8531, Japan}


\date{\today}

\begin{abstract}
Bulk-sensitive high-resolution photoemission was carried out on 
a prototype 3$d^1$ metallic Sr$_{1-x}$Ca$_x$VO$_3$. 
In a strong contrast to so far reported results, 
the bulk spectral functions are revealed to be insensitive to $x$. 
The conservation of the density of states at the Fermi level 
in spite of the electron correlation is clarified by the successful 
suppression and deconvolution of the surface contribution. 
Our study has demonstrated the importance of high-energy and 
high-resolution photoemission spectroscopy for revealing detailed bulk 
electronic states of transition metal oxides. 
\end{abstract}


\maketitle

The effect of the electron correlation to electronic states of 
transition metal (TM) oxides is still one of the most important 
and essential topics in condensed matter physics. 
The electronic structures of insulating TM oxides are basically 
understood in a framework of the Hubbard model~\cite{Hubbard} 
and/or Mott insulator description.~\cite{Mott} 
The occupied (unoccupied) $d$ orbitals hybridized 
with the neighboring O $2p$ orbitals form the 
lower (upper) Hubbard band in the presence of the electron correlation. 
The electron correlation is scaled by the on-site Coulomb 
repulsion between the $d$ electrons $U$ over a transfer integral $t$ 
between the mutually neighboring TM $3d$ states 
via the O $2p$ orbitals, namely, $U/t$.
In metallic but correlated TM oxides, 
it has been recognized that the lower and upper Hubbard bands 
remain as the so-called incoherent parts 
reflecting the electron correlation. 
A renormalized band crossing $E_F$ (coherent part) is located 
between the occupied and unoccupied incoherent parts.~\cite{AFSCVO} 
Here reported Sr$_{1-x}$Ca$_x$VO$_3$ is one of the metallic 
Mott-Hubbard systems. 
The occupied $3d$ electronic structures are expected to be simple 
because of the nominal $3d^1$ (V$^{4+}$) 
configuration. SrVO$_3$ has a cubic structure 
with the V-O-V bond angle of 180$^{\circ}$ whereas 
the angle is about 160$^{\circ}$ in CaVO$_3$ due to the smaller ionic 
radius of Ca$^{2+}$ than that of Sr$^{2+}$.~\cite{IHI98} 
It is generally thought that $t$ 
is smaller in CaVO$_3$ than in SrVO$_3$ due to the V-O-V distortion 
whereas $U$ is fairly independent of $x$. 
Therefore one might expect that $U/t$ is larger for CaVO$_3$ 
than for SrVO$_3$. 
However, measurements of the electron specific heat and magnetic susceptibility suggest that the mass-enhancement does not change 
appreciably with $x$,~\cite{IHI98} 
casting a serious question on the above scenario. 
On the other hand, accumulated theoretical studies have 
shown that the density of states (DOS) at $E_F$ is conserved 
irrespective of the presence of the electron correlation 
in the Fermi liquid if momentum dependence of the self-energy is 
negligible.~\cite{Khurana,Zhang,Bulla} 
This theorem has been proposed for a long time, 
but not experimentally confirmed yet, 
as far as our knowledge is concerned. 

Photoemission spectroscopy (PES) can directly probe occupied electronic 
structures of solids reflecting DOS. 
Valence-band PES studies of Sr$_{1-x}$Ca$_x$VO$_3$ so far 
performed at $h\nu <$ 120 eV have shown that 
the intensity at $E_F$ as well as the relative spectral weight 
of the coherent part to the incoherent part is 
systematically suppressed with increasing $x$.~\cite{IHI95,Morikawa} 
It has been believed that such a behavior originates 
from the change of $U/t$ with $x$. 
The change of the spectral weight at $E_F$ 
by the effect of $U/t$ suggests an essential role of a 
momentum-dependent self-energy, which is derived from the strong electron 
correlation.~\cite{AFSCVO,IHI95,Morikawa,AFSV} 
However, these PES spectra are very sensitive to 
the surface electronic states due to a short 
photoelectron mean free path $\lambda$.~\cite{PIS,Tanuma} 
The high-resolution valence-band PES at high-$h\nu$ near 1000 eV 
is a more direct and important technique to reveal the bulk electronic 
states near $E_F$ owing to its longer $\lambda$ of 
photoelectrons.~\cite{ASN,RSI,MaitiPRL,MaitiEU} 
In this Letter we demonstrate that genuine bulk spectral functions 
revealed by virtue of the high-energy and high-resolution PES 
for Sr$_{1-x}$Ca$_x$VO$_3$ 
on fractured surfaces are nearly independent of the Ca concentration $x$. 
This fact has been overlooked for a long time 
in low energy photoemission by the contribution of surface 
states which strongly changes with $x$. 

The PES at $h\nu =$ 900 and 275 eV was performed at 
BL25SU in SPring-8,~\cite{RSI} 
where the PES spectra were measured 
by using a GAMMADATA-SCIENTA SES-200 spectrometer. 
Single crystals of SrVO$_3$ and Sr$_{0.5}$Ca$_{0.5}$VO$_3$, 
and polycrystalline CaVO$_3$ were employed for the measurements. 
The overall energy resolution was set to about 140 and 80 meV 
at $h\nu =$ 900 and 275 eV, respectively. 
The results were compared with the low-energy PES spectra 
taken at $h\nu =$ 40.8 and 21.2 eV by using a He discharge lamp, 
which were measured by using a VG CLAM4 spectrometer in Osaka 
University. 
The energy resolution was set to 50-80 meV. The samples were cooled 
to 20 K for all the measurements. 
Clean surfaces were obtained by fracturing the samples {\it in situ} 
at measuring temperatures 
and the surface cleanliness was confirmed before and 
after the measurements. The base pressure was about 4 x 10$^{-8}$ Pa. 

\begin{figure}
\includegraphics[width=8.5cm,clip]{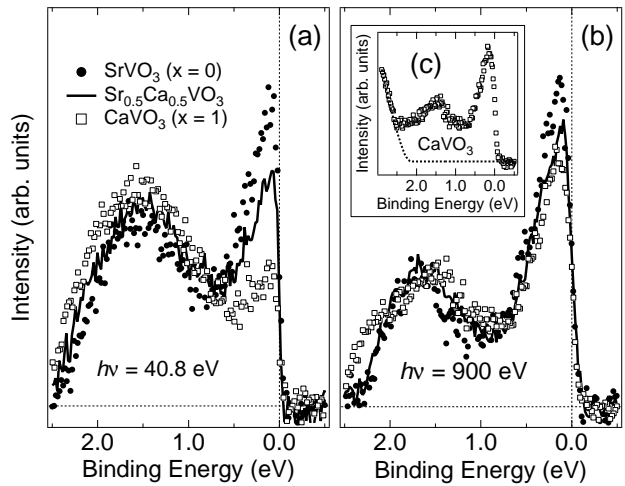}%
\caption{\label{Fig1}High-resolution V $3d$ spectral weights near $E_F$ 
of Sr$_{1-x}$Ca$_x$VO$_3$ at (a) low-energy $h\nu =$ 40.8 
and (b) high-energy $h\nu =$ 900 eV, 
which have been obtained by subtracting the fitted tails of O $2p$ 
contributions from the raw spectra (dashed line and squares in 
the inset (c), respectively).}
\end{figure}

\begin{figure}
\includegraphics[width=6cm,clip]{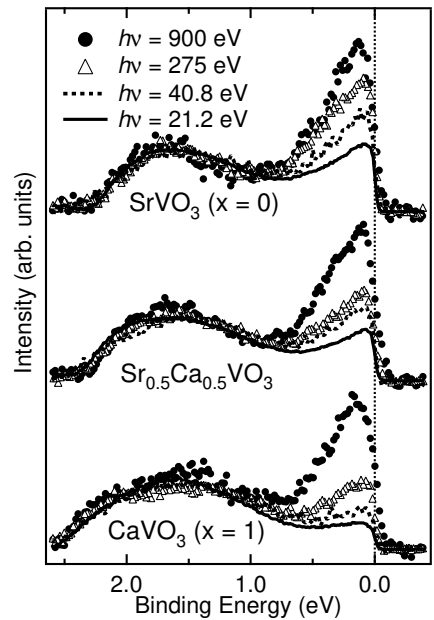}%
\caption{\label{Fig2}Photon-energy dependence of the V $3d$ spectral 
weights for Sr$_{1-x}$Ca$_x$VO$_3$. 
The V $3d$ spectra are normalized by the integrated intensities of 
the incoherent part ranging from 0.8 to 2.6 eV.}
\end{figure}

Figures~\ref{Fig1} and \ref{Fig2} show the high-resolution PES 
spectra near $E_F$ measured at low- and high-energy excitations. 
In all the spectra, the peak near $E_F$ and 
the broad peak centered at about 1.6 eV are corresponding to 
the coherent and incoherent parts, respectively. 
These structures are drastically enhanced in a V $2p-3d$ resonance PES 
(not shown) and therefore originate predominantly 
from the V $3d$ electronic states.  
On going from CaVO$_3$ to SrVO$_3$, 
the coherent part is clearly enhanced in the low-energy PES spectra 
at $h\nu =$ 40.8 eV as shown in Fig.~\ref{Fig1}(a). 
This tendency is consistent with the previous low-energy PES 
studies.~\cite{IHI95,Morikawa} 
However, this spectral difference among the compounds becomes 
noticeably smaller in the spectra measured at 
$h\nu =$ 900 eV in Fig.~\ref{Fig1}(b). 
The photon-energy dependence of the V $3d$ spectral weights 
for Sr$_{1-x}$Ca$_x$VO$_3$ is summarized in Fig.~\ref{Fig2}. 
The coherent spectral weight increases drastically with the 
photon energy for all the compounds. 
One might consider that the relatively strong incoherent spectral 
weight of the 1.6 eV peak at low-$h\nu$ originates possibly from a 
large O $2p$ weight in this peak because the low-energy PES of 
TM oxides is generally more sensitive to the O $2p$ states 
than the TM $3d$ states. 
However, the observed behavior with $h\nu$ is rather independent of 
the relative photoionization cross section of the V $3d$ to O $2p$ 
states,~\cite{Lindau} 
which should be the largest at 275 eV among $h\nu =$ 21.2, 40.8, 
275 and 900 eV. 
Therefore the monotonous increase of the coherent part with $h\nu$ 
should originate from the increased sensitivity to the bulk $3d$ 
states at higher $h\nu$. 
Namely, the V $3d$ spectra at low-$h\nu$ mainly reflect the 
surface electronic states which are more localized than the bulk states 
caused by the broken V-O-V topological connectivity and/or 
structural surface relaxations at the surface. 

We have estimated the bulk and surface V $3d$ spectral functions 
of Sr$_{1-x}$Ca$_x$VO$_3$ from the experimental spectra at $h\nu =$ 
900 and 275 eV by the following procedure: 
(1) The mean free path $\lambda$ has been calculated as $\sim$17 and 
$\sim$7 {\AA} at $h\nu =$ 900 and 275 eV.~\cite{Tanuma} 
(2) The bulk weight $R$ ($< 1$, depending on $h\nu$) should be 
determined as $exp(-s/\lambda )$ where $s$ is a "surface thickness", 
therefore $R$'s at 900 eV ($R_{900}$) and 275 eV ($R_{275}$) 
are related as $R_{275} = R_{900}^{2.4}$. 
(3) The observed V $3d$ spectrum 
$I(E_B)$ at $h\nu =$ 900 eV is represented as 
$I_{900}(E_B) = bulk(E_B)R_{900} + surf(E_B)(1-R_{900})$ 
while 
$I_{275}(E_B) = bulk(E_B)R_{900}^{2.4} + surf(E_B)(1-R_{900}^{2.4})$, 
where $bulk(E_B)$ [$surf(E_B)$] is the bulk (surface) $3d$ spectral 
function and $E_B$ stands for the electron binding energy. 
(4) If $s$ is assumed as 7.5 {\AA} corresponding to about twice the 
V-O-V distance,~\cite{IHI98} $R_{900}$ ($R_{275}$) is determined as 
$\sim$0.64 ($\sim$0.34). 
The bulk and surface $3d$ spectral functions $bulk(E_B)$ and 
$surf(E_B)$ are consequently obtained from the observed spectra 
measured at $h\nu =$ 900 and 275 eV. 
It should be noted that the line shapes of $bulk(E_B)$ have hardly 
changed even when we assume $s$ as from 5.4 to 11 \AA. 
If $s$ is assumed to be less than 5.4 \AA, 
unrealistic "negative" intensities are seen in some 
binding energy region for $surf(E_B)$.
The uncertainty of $s$ (5.4-11 \AA) yields the error for $R_{900}$ 
and $R_{275}$ as $\pm$0.12, by which the error of the 
intensity in the vicinity of $E_F$ in the bulk spectra is estimated 
to be less than 15 \%. 
The results for the bulk states are shown in Fig. 3(a). 
In contrast to the previous PES studies, the bulk $3d$ spectral 
functions are almost equivalent among the three compounds 
and the intensity in the vicinity of $E_F$ does not change 
for different $x$. 
These results indicate that the effect of the V-O-V distortion is not 
influential in the bulk $3d$ electronic states in Sr$_{1-x}$Ca$_x$VO$_3$. 
Our bulk-sensitive results are consistent with the behavior of 
the electron specific heat and magnetic susceptibility measurements. 
This fact has never been revealed by so far reported low-energy PES. 
It should be even noticed that the bulk spectral functions nearly 
independent of $x$ could not be obtained 
on scraped sample surfaces.~\cite{MaitiEU} 

\begin{figure}
\includegraphics[width=6cm,clip]{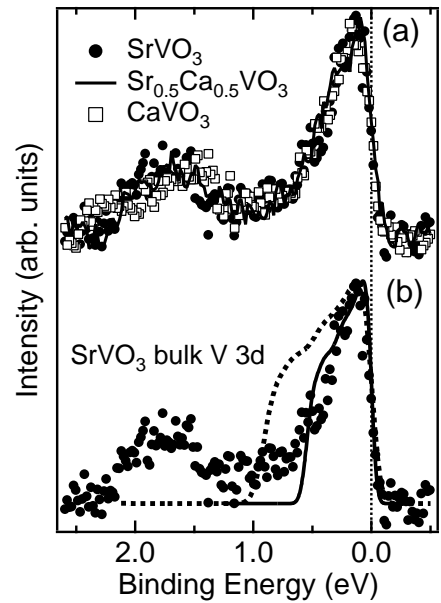}%
\caption{\label{Fig3}(a) Bulk V $3d$ spectral functions of SrVO$_3$ 
(closed circles), Sr$_{0.5}$Ca$_{0.5}$VO$_3$ (solid line) and CaVO$_3$ 
(open squares). 
(b) Comparison of the experimentally obtained bulk V $3d$ spectral 
function of SrVO$_3$ (closed circles) to the V $3d$ partial density 
of states for SrVO$_3$ (dashed curve) obtained 
from the band-structure calculation, 
which has been broadened by the experimental resolution of 140 meV. 
The solid curve shows the same V $3d$ partial density of states 
but the energy is scaled down by a factor of 0.6.}
\end{figure}

In order to know the genuine effect of the electron correlation 
in the bulk for this system, 
we compare the bulk spectrum with a V $3d$ partial DOS for SrVO$_3$ 
obtained from a band-structure calculation~\cite{FLAPW} 
as shown in Fig.~\ref{Fig3}(b). 
Here, the partial DOS broadened by the instrumental resolution 
(dashed curve) is normalized to the bulk $3d$ spectral 
function of SrVO$_3$ by the integrated intensity from $E_F$ to 2.6 eV. 
This comparison shows us two obvious discrepancies. 
At first, the band-structure calculation can not reproduce 
the incoherent spectral weight centered at 1.6 eV at all. 
Secondly, the width of the observed coherent part is about 
60 \% of the predicted value. 
These features indicate that the vanishing coherent spectral 
weight due to the narrowing is transferred to the incoherent part. 
Our study has revealed that the bulk electron correlation effect, 
which can not be fully taken into account in the band-structure 
calculation,~\cite{AFSCVO,Zhang,Bulla,IHI95,Morikawa,AFSV} 
is strongly modifying the predicted bulk spectral function 
resulting in the spectral narrowing and redistribution. 
Nonetheless, it is found that the band-structure calculation really 
well reproduces the spectral intensity in the vicinity of $E_F$. 
This result provides the first experimental evidence 
for the conservation of DOS at $E_F$ against the electron 
correlation,~\cite{T0} 
suggesting that the self-energy has nearly no momentum dependence in 
Sr$_{1-x}$Ca$_x$VO$_3$.~\cite{DOSCVO}  
Thus such a treatment as a dynamical mean-field theory 
(DMFT),~\cite{Bulla} in which the momentum dependence of the 
self-energy is not taken into account, can be a useful approach. 

\begin{figure}
\includegraphics[width=5cm,clip]{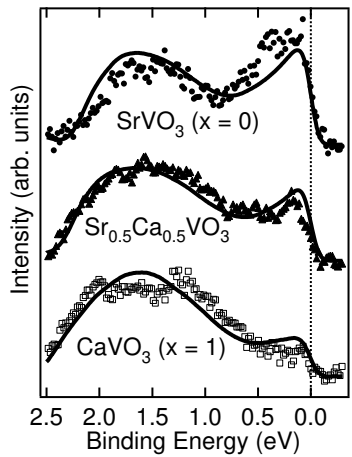}%
\caption{\label{Fig4}Comparison of the surface $3d$ spectral functions 
(dots) estimated from our data by assuming the surface thickness 
$s$ of 7.5 {\AA} with the $3d$ spectral weights 
measured at $h\nu =$ 21.2 eV (solid curves) broadened by 
the experimental resolution at $h\nu =$ 900 eV ($\sim$140 meV).}
\end{figure}

In Fig.~\ref{Fig4} are summarized the estimated surface $3d$ spectral 
functions by symbols. 
We notice that the coherent part near $E_F$ increases on going from 
CaVO$_3$ to SrVO$_3$. 
Figure~\ref{Fig4} also demonstrates that 
these spectra coincide well with the broadened $3d$ spectra 
measured at $h\nu =$ 21.2 eV as shown by solid curves, 
verifying that the surface contribution is predominant 
in the $h\nu =$ 21.2 eV spectra.~\cite{BulkHe2} 
Some people believe that $\lambda$ generally takes a 
minimum in the kinetic energy region of $\sim$100 eV 
and becomes longer again at low kinetic energies, e.g. $\sim$20 eV. 
Such an energy dependence has been known as a 
"universal curve".~\cite{PIS} 
On the other hand, Figs.~\ref{Fig2} and \ref{Fig4} suggest that 
the mean free path of the V $3d$ photoelectrons decreases with 
kinetic energy down to $\sim$20 eV 
even in the low kinetic energy region. 
In this respect, a previous photoemission study of rare earth 
metals has indicated that $\lambda$ of $4f$ photoelectrons 
decreases also with the kinetic energy down to 
$\sim$20 eV,~\cite{Gerken85} 
being again completely inconsistent with the so-called universal curve. 
The surface-sensitivity of photoemission spectroscopy depends upon 
the individual orbital of individual material at low-$h\nu$. 
Therefore bulk electronic structures should be re-examined 
by high-resolution high-energy photoemission spectroscopy 
especially for strongly correlated $3d$ and $4f$ electrons systems. 

In summary, our results have demonstrated for the first time 
the importance of the high-energy and high-resolution photoemission 
spectroscopy for revealing the genuine 
bulk electronic states of the correlated $3d$ TM oxides. 
We have confirmed that the intensity of the spectral function at $E_F$ 
is not much changed in the bulk spectra even when the 
electron correlation is effective in the TM oxides. 
Future theoretical studies will clarify why the surface $3d$ 
electronic states change appreciably with $x$ 
whereas the bulk states are insensitive to $x$. 

We thank T. Ushida, A. Shigemoto, T. Satonaka, T. Iwasaki, M. Okazaki, 
S. Kasai, A. Higashiya, K. Konoike, A. Yamasaki, A. Irizawa 
and the staff of SPring-8, especially T. Muro, T. Matsushita 
and T. Nakatani for supporting the experiments. 
This work was supported by a Grant-in-Aid for COE Research 
from the Ministry of Education, Culture, Sports, Science 
and Technology (MEXT), Japan. 
The photoemission at $h\nu =$ 275 and 900 eV was performed 
under the approval of the Japan Synchrotron Radiation Research 
Institute (Proposal No. 1999B0076-NS-np).


\references
\bibitem{Hubbard}J. Hubbard, Proc. Roy. Soc. A {\bf 276}, 238 (1963).
\bibitem{Mott}N. F. Mott, Metal-insulator transitions. 
Second Edition (Taylor and Francis, London, 1990).
\bibitem{AFSCVO}A. Fujimori {\it et al.}, Phys. Rev. Lett. 
{\bf 69}, 1796 (1992).
\bibitem{IHI98}I. H. Inoue {\it et al.}, Phys. Rev. B {\bf 58}, 4372 
(1998).
\bibitem{Khurana}A. Khurana, Phys. Rev. B {\bf 40}, 4316 (1989).
\bibitem{Zhang}X. Y. Zhang, M. J. Rozenberg, and G. Kotliar, 
Phys. Rev. Lett. {\bf 70}, 1666 (1993).
\bibitem{Bulla}R. Bulla, T. A. Costi, and D. Vollhardt, 
Phys. Rev. B {\bf 64}, 045103 (2001).
\bibitem{IHI95}I. H. Inoue {\it et al.}, Phys. Rev. Lett. {\bf 74}, 
2539 (1995).
\bibitem{Morikawa}K. Morikawa {\it et al.}, Phys. Rev. B {\bf 52}, 
13711 (1995).
\bibitem{AFSV}A. Fujimori {\it et al.}, Spectroscopy of Mott Insulators 
and Correlated Metals, A. Fujimori, Y. Tokura, Eds. 
(Springer-Verlag, Berlin, 1995), p. 174.
\bibitem{PIS}D. A. Shirley, Photoemission in Solids I, M. Cardona, 
L. Ley, Eds. (Springer-Verlag, Berlin, 1978).
\bibitem{Tanuma}S. Tanuma, C. J. Powell, and D. R. Penn, 
Surf. Sci. {\bf 192}, L849 (1987).
\bibitem{ASN}A. Sekiyama {\it et al.}, Nature {\bf 403}, 396 (2000).
\bibitem{RSI}Y. Saitoh {\it et al.}, Rev. Sci. Instrum. {\bf 71}, 
3254 (2000).
\bibitem{MaitiPRL}K. Maiti, P. Mahadevan, and D. D. Sarma, 
Phys. Rev. Lett. {\bf 80}, 2885 (1998).
\bibitem{MaitiEU}K. Maiti {\it et al.}, Europhys. Lett. {\bf 55}, 
246 (2001). 
\bibitem{Lindau}J. J. Yeh and I. Lindau, At. Data Nucl. Data Tables 
{\bf 32}, 1 (1985).
\bibitem{FLAPW}The band-structure calculation performed 
by using a full-potential linearized augmented plane wave (FLAPW) 
method within the local density approximation. This method is one of 
the most reliable methods in the band-structure calculations.
\bibitem{T0}This theorem is applicable only at zero temperature. 
According to the finite temperature DMFT,~\cite{Bulla} however, 
an effect of the finite temperature T = 20 K (0.0017 eV) in our 
measurements is negligible compared with the band width $W$ 
($\agt$1 eV) when bulk SrVO$_3$ is far from the Mott transition. 
\bibitem{DOSCVO}According to the band-structure calculation in 
Ref.~\cite{IHI95}, the V $3d$ partial DOS of CaVO$_3$ is 
very similar to that of SrVO$_3$. 
Therefore the conservation of DOS at $E_F$ should also be 
applicable to CaVO$_3$ from our results. 
\bibitem{BulkHe2}The surface contributions in the spectra measured 
at $h\nu$ = 40.8 eV are estimated as at least 70 \% from our results. 
\bibitem{Gerken85}F. Gerken {\it et al.}, Physica Scripta {\bf 32}, 
43 (1985).
\end{document}